\documentclass[]{spie}  

\usepackage[]{graphicx}

\title{Long-term trends in radiation damage of \\ Chandra X-ray CCDs} 

\author{C. E. Grant, M. W. Bautz, S. M. Kissel, B. LaMarr, G. Y. Prigozhin
\skiplinehalf
Kavli Institute for Astrophysics and Space Research, \\
Massachusetts Institute of Technology \\
Cambridge, Massachusetts 02139
}

\authorinfo{Further author information: (Send correspondence to C.E.G.)\\C.E.G.: E-mail: cgrant@space.mit.edu\\Copyright 2005 Society of Photo-Optical Instrumentation Engineers.\\This paper will be published in Proc. SPIE, vol. 5898, and is made available as an electronic reprint with permission of SPIE.  One print or electronic copy may be made for personal use only.  Systematic or multiple reproduction of any material in this paper for a fee or for commercial purposes, or modification of the content of the paper are prohibited.}

\begin{document} 
  \maketitle 

\begin{abstract}
Soon after launch, the Advanced CCD Imaging Spectrometer (ACIS), one of the focal plane instruments on the Chandra X-ray Observatory, suffered radiation damage from exposure to soft protons during passages through the Earth's radiation belts.  Current operations require ACIS to be protected during radiation belt passages to prevent this type of damage, but there remains a much slower and more gradual increase.  We present the history of ACIS charge transfer inefficiency (CTI), and other measures of radiation damage, from January 2000 through June 2005.  The rate of CTI increase is low, of order $10^{-6}$ per year, with no indication of step-function increases due to specific solar events.  Based on the time history and CCD location of the CTI increase, we speculate on the nature of the damaging particles.
\end{abstract}

\keywords{Charge Coupled Devices, radiation damage, charge transfer inefficiency, Chandra X-ray Observatory, ACIS}

\section{INTRODUCTION}
\label{sect:intro}
The Chandra X-ray Observatory, the third of NASA's great observatories in space, was launched just past midnight on July 23, 1999, aboard the space shuttle {\it Columbia}\cite{cha2}.  After a series of orbital maneuvers Chandra reached its final, highly elliptical, orbit.  Chandra's orbit, with a perigee of 10,000~km, an apogee of 140,000~km and an initial inclination of 28.5$^\circ$, transits a wide range of particle environments, from the radiation belts at closest approach through the magnetosphere and magnetopause and past the bow shock into the solar wind.

The Advanced CCD Imaging Spectrometer (ACIS), one of two focal plane science instruments on Chandra, utilizes frame-transfer charge-coupled devices (CCDs) of two types, front- and back-illuminated (FI and BI).  Soon after launch it was discovered that the FI CCDs had suffered radiation damage from exposure to soft protons scattered off the Observatory's grazing-incidence optics during passages through the Earth's radiation belts\cite{gyp00}.  Since mid-September 1999, ACIS has been protected during radiation belt passages and there is an ongoing effort to prevent further damage and to develop hardware and software strategies to mitigate the effects of charge transfer inefficiency on data analysis\cite{cticorr}.  Ref.~\citenum{odell} discusses the radiation management strategy on Chandra.

This paper begins by describing the calibration data used to monitor radiation damage in Section~\ref{sect:data}.  Section~\ref{sect:raddamage} describes in detail the characteristics of ACIS radiation damage.  The evolution of ACIS radiation damage is discussed in Section~\ref{sect:evol}.  Section~\ref{sect:conc} speculates on the nature of the on-going damage.

\section{DATA}
\label{sect:data}

\begin{figure}
\vspace{3.4in}
\includegraphics{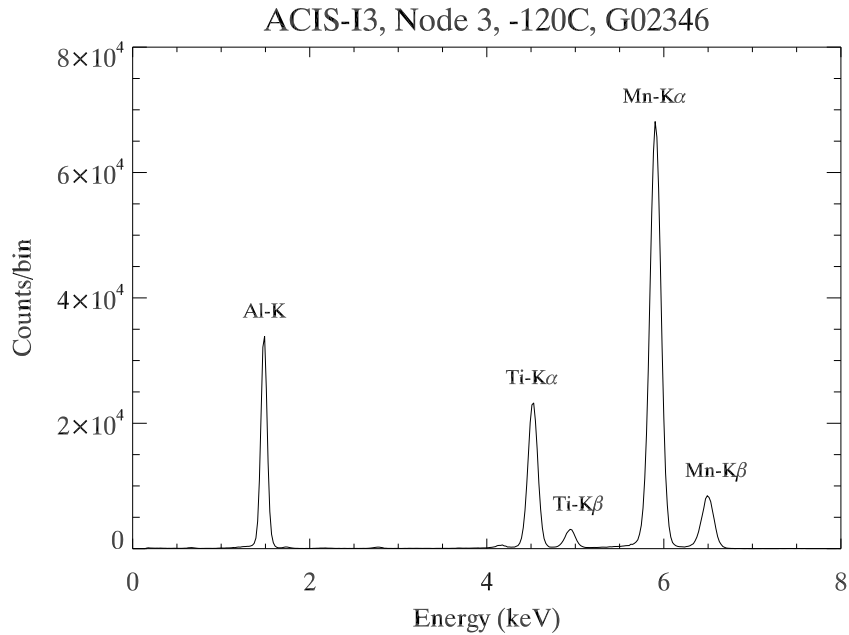}
\caption{The spectrum of the ACIS calibration source on the ACIS-I3 FI CCD.  Only data from the first 256 rows are included to minimize CTI degradation.}
\label{fig:ecsspectrum}
\end{figure}

All the results shown here are based on data taken of the ACIS External Calibration Source (ECS) which uniformly illuminates the ACIS CCDs when ACIS is moved out of the telescope focus.  Since the discovery of the initial radiation damage, a continuing series of observations of the ECS have been undertaken just before and after the instruments are safed for perigee passage to monitor the performance of the ACIS CCDs.  ACIS is placed in the stowed position exposing the CCDs to the ECS which produces many spectral features, the strongest of which are Mn-K$\alpha$ (5.9~keV), Ti-K$\alpha$ (4.5~keV), and Al-K (1.5~keV).  Figure~\ref{fig:ecsspectrum} shows a spectrum of the ECS and labels the brightest features.  The data are taken in the standard Timed Exposure mode with a 3.2~second frame time.  Typical exposure times for each observation range from 5.5 to 8~ksecs.  ACIS events are recorded as 3 x 3 pixel event islands.  All the observations used in this paper were taken with a focal plane temperature set-point of --120$^\circ$~C and all the analysis concentrates on the Mn-K$\alpha$ spectral line at 5.9~keV.

\section{CHARACTERISTICS OF ACIS RADIATION DAMAGE} 
\label{sect:raddamage}

One symptom of radiation damage in CCDs is an increase in the number of charge traps.  When charge is transfered across the CCD to the readout, some portion can be captured by the traps and gradually re-emitted.  If the original charge packet has been transfered away before the traps re-emit, the captured charge is ``lost'' to the charge packet. The pulseheight read out from the instrument which corresponds to a given energy decreases with increasing transfer distance.  This process is quantified as charge transfer inefficiency (CTI), the fractional charge loss per pixel and is calculated from a linear fit to the pulseheight versus row number; CTI = (slope/intercept).  X-ray events on ACIS CCDs can produce charge packets which occupy multiple pixels.  We, however, compute CTI from the center pixel pulseheight alone.

Damage can exist in the imaging or framestore array, causing parallel CTI in the column direction, or in the serial register, causing serial CTI along rows.  The ACIS CCDs come in two flavors, front- and back-illuminated, which have different manifestations of CTI.  The distribution of re-emission time constants of the electron traps which cause CTI vary depending on the type of damage.  In addition, the pixel to pixel transfer time in the imaging, framestore and serial arrays differs so that the same species of electron trap can produce different CTI results.

Measured CTI is a function of fluence or, more specifically, the amount of charge deposited on the CCD.\cite{gendreau}  As the fluence increases, traps filled by one charge packet may remain filled as a second charge packet is transferred through the pixel.  The second charge packet sees fewer unoccupied traps as a result of the previous ``sacrificial charge'' and loses less charge then it would have otherwise.  Whether sacrificial charge is important depends on the interaction of the pixel transfer time, the typical distance between charge packets and the re-emission time constant of the traps.  For observations of the ACIS ECS and for most Chandra observations in general, the primary source of sacrificial charge is the particle background.  The measured CTI will therefore be a function of the particle environment. (For more on sacrificial charge in ACIS CCDs, see Ref.~\citenum{saccharge})

In addition to sacrificial charge effects, measured CTI is also a function of temperature.  The detrapping time constant decreases as the temperature increases so that different populations of traps can become more or less important.  The distribution of trap time constants at a particular temperature determines the CTI, so temperature can positively or negatively correlate with CTI.  The standard operating temperature of the ACIS focal plane is --120$^\circ$~C. In some spacecraft orientations, such as when the bright Earth is visible to the ACIS radiator, the cooling efficiency is reduced such that the focal plane temperature rises occasionally by as much as five degrees.  This is most common during perigee passages so that ECS measurements have much larger temperature variations than do typical science observations.

There are a number of ways in which CTI changes the overall instrument performance.  The cumulative charge loss causes a strong position dependence in the pulseheight, the line response function and the quantum efficiency.  At the bottom of the CCD closest to the readout the performance is the same as the original undamaged CCD.  The pulseheight of each spectral line decreases with increasing row number due to the increasing transfer distance.  The width of each spectral feature also increases with increasing transfer distance due to a number of factors.  The charge trapping and re-emission process is stochastic, and this additional noise in the pulseheight distribution increases with each transfer.  The trap distribution is non-uniform, so without calibration on the scale of a few pixels, this adds to the spectral width.  Finally, there are variations in trap occupancy which increase the charge loss distribution\cite{saccharge}.  All of these effects are applied to the charge in each pixel, so multi-pixel events will be more degraded than single-pixel events.

The charge loss and re-emission process can also change the morphology of the X-ray event.  ACIS events are recorded as 3 x 3 pixel islands.  Each event island is assigned a grade based on the charge distribution.  In this manner X-ray events can be distinguished from the much more common cosmic ray events which have distinctly different event island distributions.  Events with grades that are least likely to be from X-rays are ignored in analysis and a subset are discarded on-board to reduce telemetry requirements.  After the cumulative charge loss and re-emission during transfer, X-ray events can be smeared into morphologies commonly associated with cosmic rays.  The produces a row-dependent loss of quantum efficiency.  At the current damage level of the ACIS FI CCDs, the quantum efficiency loss is small and is only important for high energy X-rays which have a higher proportion of multi-pixel events.

Measuring parallel CTI is relatively straight-forward; simply fit a linear function to the pulseheight as a function of row.  Serial CTI is measured in much the same way, using column instead of row number, but can be difficult to disentangle from parallel CTI effects.  We measure serial CTI by selecting events from the bottom rows of the CCD, thus minimizing any crosstalk with parallel CTI.  Since no X-ray events occur in the framestore, it is difficult to directly measure parallel CTI in the framestore array.  If significant charge loss occurs in the framestore, we expect to see a non-linear flattening of charge loss at the bottom of the imaging array due to the change in sacrificial charge environment from the framestore to imaging arrays.  It is possible, however, to estimate changes in framestore CTI by monitoring the pulseheight of events at the bottom of the imaging array.  Assuming that the electronic gain remains the same, any pulseheight decrease can be used to estimate the change in framestore CTI.

\begin{figure}
\vspace{3.4in}
\includegraphics{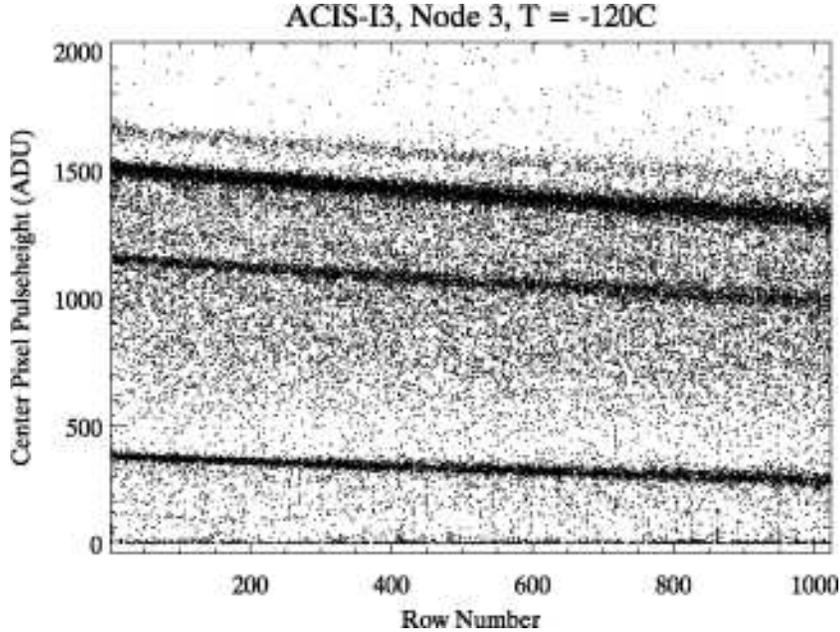}
\caption{A scatter plot of the center pixel pulseheight of each X-ray event versus its row number for a single node.  The diagonal lines are spectral features in the ACIS calibration source.  Events split over multiple pixels appear at lower pulseheights than equivalent single pixel events.}
\label{fig:phvrow}
\end{figure}

\subsection{CTI in Front-illuminated CCDs}

The eight front-illuminated CCDs had essentially no CTI before launch, but are strongly sensitive to radiation damage from low energy protons ($\sim$100~keV) which preferentially create traps in the buried transfer channel.  The framestore covers are thick enough to stop this radiation, so the initial damage was limited to the imaging area of the FI CCDs.  Radiation damage from low-energy protons is now minimized by moving the ACIS detector away from the aimpoint of the observatory during passages through the Earth's particle belts.  Continuing exposure to both low and high energy particles over the lifetime of the mission slowly degrades the CTI further.  Figure~\ref{fig:phvrow} is a scatter plot of the center pixel pulseheight of each X-ray event versus its row number for a single node of the FI CCD I3.  Each of the diagonal lines is a spectral feature in the ACIS calibration source.  The pulseheight of each spectral line decreases with increasing row number as the line width increases. As of January 2000, the parallel CTI at 5.9~keV of the ACIS FI CCDs varied across the focal plane from $1 - 2 \times 10^{-4}$ at the nominal operating temperature of --120$^\circ$~C.  Parallel CTI in the framestore array and serial CTI were not affected by the initial radiation damage and remain negligible with upper limits of $< 10^{-6}$ (framestore array) and $< 2 \times 10^{-5}$ (serial array).

\subsection{CTI in Back-illuminated CCDs}

The two back-illuminated CCDs (ACIS-S1,S3) suffered damage during the manufacturing process and exhibit CTI in both the imaging and framestore areas and the serial transfer array, but are less sensitive to the low energy particles which damage the FI CCDs because they cannot reach the transfer channel.  The parallel CTI at 5.9~keV of the ACIS-S3 BI CCD was  $\sim 1.6 \times 10^{-5}$ at a temperature of --120$^\circ$~C at the beginning of the mission with a strong non-linear flattening of pulseheight at low row numbers due to CTI in the framestore array.  The serial CTI is much larger, $\sim 8 \times 10^{-5}$.  The ACIS-S1 device has poorer performance than the S3, and the parallel and serial CTI are approximately twice that of S3.

{\subsection{Trailing Charge} \label{sec:trail}}

We also use an additional metric, the trailing charge fraction, to characterize the radiation damage on ACIS CCDs.  CTI is a measure of the quantity of charge traps, but does not directly characterize the distribution of trap time constants, except indirectly through the interaction with sacrificial charge.  A better accounting of the trap time constants can be done with careful analysis of the trailing charge behind each event.  As the events are transfered the charge lost to the original packet is re-emitted into the following pixels.  The exponential decay in the trailing charge can be used to determine the time constants of the electron traps.  Ref. \citenum{gyp00} describes such an analysis for the ACIS CCDs using full-frame data.  Telemetry constraints limit the quantity of full-frame data available, so we concentrate here on the charge in the first trailing pixel which is always telemetered as part of the standard 3 x 3 pixel event island.  The ratio of the charge in the first trailing pixel to the charge lost from the central pixel can be used to characterize the trap distribution in a general way.  A higher trailing charge fraction implies more traps with shorter time constants which deposit most of their captured charge close to the original event.  The trailing charge fraction for the FI CCDs is $\sim 0.04$, while for the BI CCDs it is closer to 0.4 (parallel) and 0.15 (serial).

\section{EVOLUTION OF ACIS RADIATION DAMAGE}
\label{sect:evol}

\begin{figure}
\vspace{2.7in}
\includegraphics{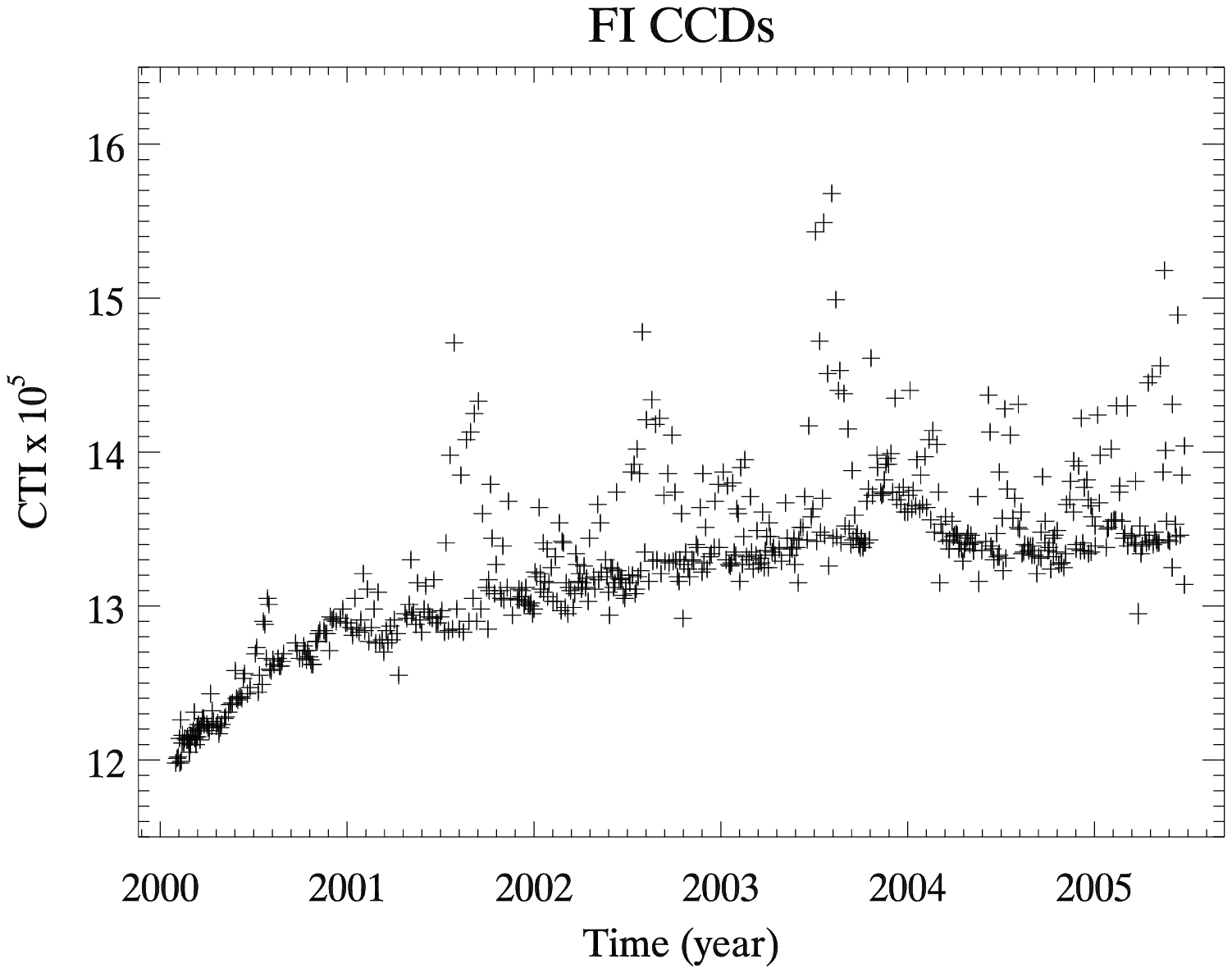}
\includegraphics{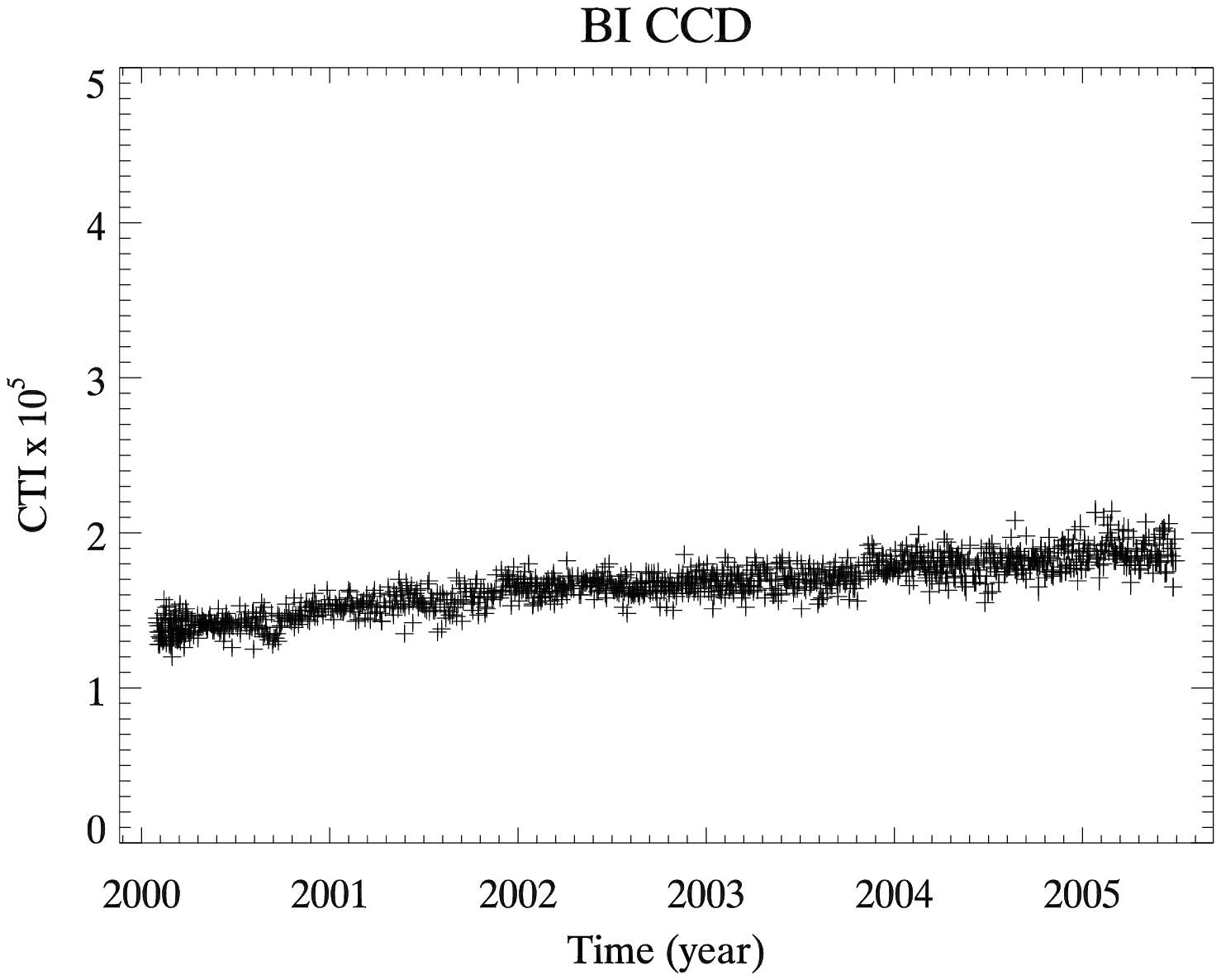}
\caption{The measured parallel CTI for the I-array FI CCDs (left) and the S3 BI CCD (right).  No corrections have been applied for focal plane temperature or particle background variations.  The upward scatter for the FI CCDs is due to warmer than nominal focal plane temperatures, while the sawtooth shapes in the lowest CTI numbers are due to sacrificial charge from the particle background.  The BI CCD is clearly less sensitive to both effects.}
\label{fig:rawcti}
\end{figure}

The characteristics of the traps in the FI CCDs are highly sensitive to the ambient particle background and the focal plane temperature, while the BI CCDs are less sensitive.  Monitoring the actual level of radiation damage requires removing both these effects.  Figure~\ref{fig:rawcti} shows the measured parallel CTI since January 2000 without any corrections for temperature or sacrificial charge from the particle background.  The mean of the four ACIS-I CCDs (I0-I3) are used to typify the behavior of FI CCDs while the ACIS-S3 CCD is used for BI CCDs.

\subsection{CTI Dependence on Particle Background}

\begin{figure}
\vspace{3.4in}
\includegraphics{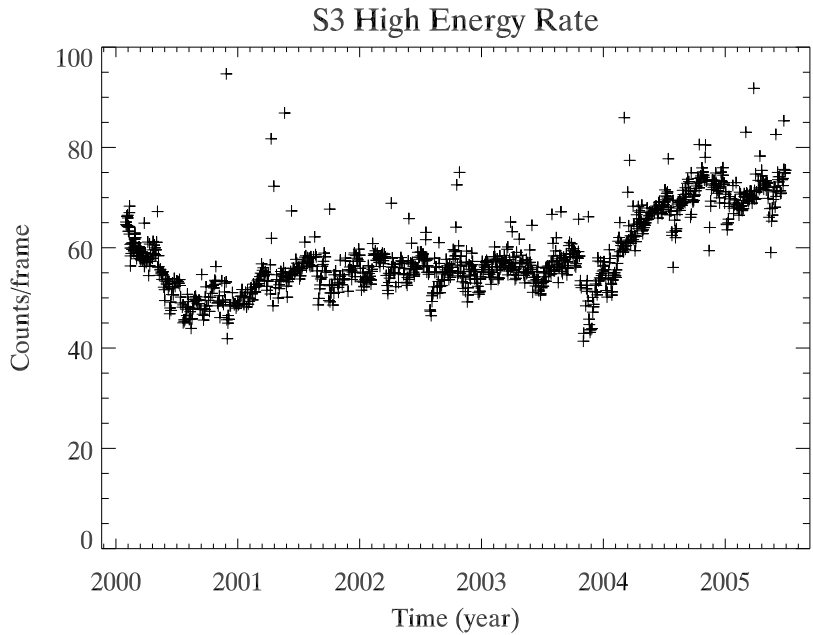}
\caption{The long term trend of the ACIS particle background as measured by the high energy rate on the ACIS-S3 BI CCD.  Each data point represents a single CTI observation.  The dip in the second half of 2000 coincides with solar maximum.  The downward jump in the last quarter of 2003 is due to the ``Halloween'' 2003 solar storm.}
\label{fig:bkg}
\end{figure}

In Chandra's high altitude orbit, cosmic rays account for a large fraction of the charge deposited on the CCDs.  If the cosmic ray background rate were changing, one would expect to see the changes reflected in the measured CTI.  The time period of this study encompasses many interesting events that affect the particle background such as nearly half a solar cycle, solar maximum, and the largest solar X-ray flare ever observed.  As a measure of the particle background rate in situ, we are using the counting rate of events on the BI CCD S3 with pulseheights greater than 3750~ADU ($\sim 15$~keV).  At these high energies, the effective area of the telescope mirrors is essentially zero, so these events are caused by energetic particles rather than astrophysical X-rays. Using S3, which is more robust against the low-energy protons that caused the initial damage, should should minimize any radiation damage induced structures.  Figure~\ref{fig:bkg} shows the long-term trend of the S3 high energy rate which correlates well with high energy ($>$ 10 MeV) protons \cite{bkg}.  Some of the structures seen in the particle background rate can also be seen in the measured CTI shown in Figure~\ref{fig:rawcti}.

\begin{figure}
\vspace{2.7in}
\includegraphics{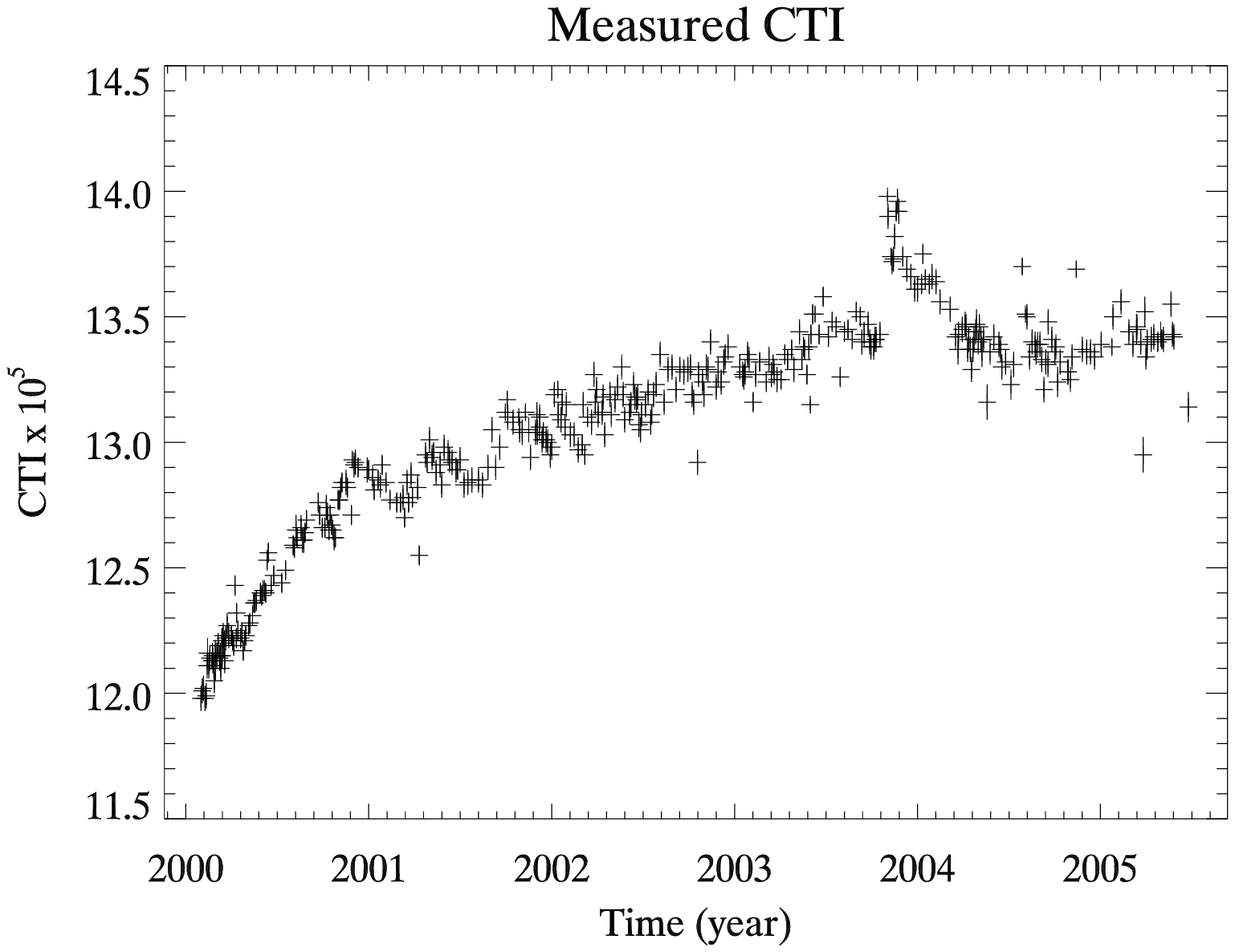}
\includegraphics{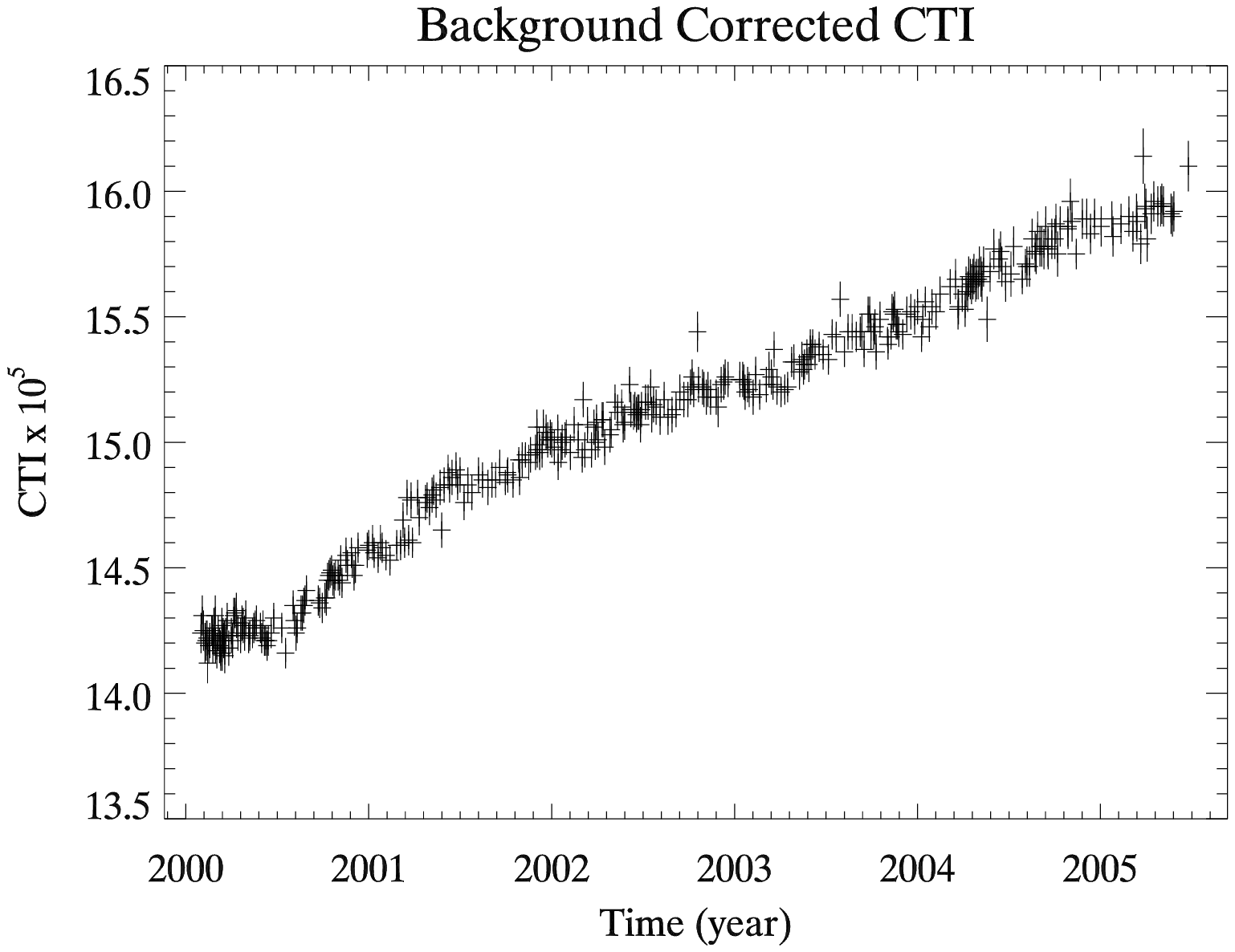}
\caption{Parallel CTI for the I-array FI CCDs before (left) and after (right) correcting for particle background variations.  The data have been filtered to remove temperature dependence.  Many of the structures seen in the uncorrected data are absent after correcting for the particle background variation.}
\label{fig:ctifix1}
\end{figure}

Initially, when the particle background was monotonically decreasing and the CTI was increasing, it was difficult to differentiate between radiation damage and sacrificial charge effects.  Now that degeneracy is broken and CTI data with identical background rates taken years apart can be used to determine the true increase due to continuing radiation damage.  We have chosen to group the CTI data in bins two counts/frame wide.  To remove the temperature variations, we only select data with a mean temperature of --119.9$^\circ$ to --119.7$^\circ$~C.  The true CTI increase can then be calculated for each group of identical background rate CTI measurements.  We can then reverse this process and determine the correction factor for the measured CTI that removes the background dependence.  Parallel CTI on the FI CCDs is relatively sensitive to changes in the particle background rate.  A change of 10\% in the background rate produces a change of $\sim 2 \times 10^{-6}$ or about 1.6\% in measured CTI.  The BI CCD parallel CTI is less sensitive -- a change of 10\% in the background rate produces a change of $\sim 2 \times 10^{-7}$ or about 1.0\% in the measured CTI.  For the BI CCD serial CTI, a change of 10\% in the background rate produces a change of $\sim 2 \times 10^{-6}$ or about 16\%.  Figure~\ref{fig:ctifix1} shows the I-array FI CCD CTI before and after correcting for particle background variations.

\subsection{CTI Dependence on Temperature}

When the ACIS door was first opened, not long after launch, the focal plane temperature was kept warmer than nominal, in order to reduce the accumulation of contamination.  The temperature was stepped downward until January 2000 when the focal plane was set to the nominal --120$^\circ$~C.  This temperature was selected to minimize performance degradation of the radiation damage that was expected throughout the mission lifetime and is the lowest temperature that can be feasible maintained.  In some spacecraft orientations, such as when the bright Earth is visible to the ACIS radiator, the cooling efficiency is reduced such that the focal plane temperature rises occasionally by as much as five degrees.  This is most common during perigee passages so that ECS measurements have much larger temperature variations than do typical science observations.  The frequency of such deviations has increased slightly, but much of that is due to the implementation of new constraints on spacecraft orientation from other systems.  The nominal cold state has not drifted upwards more than 0.2$^\circ$~C.

\begin{figure}
\vspace{2.7in}
\includegraphics{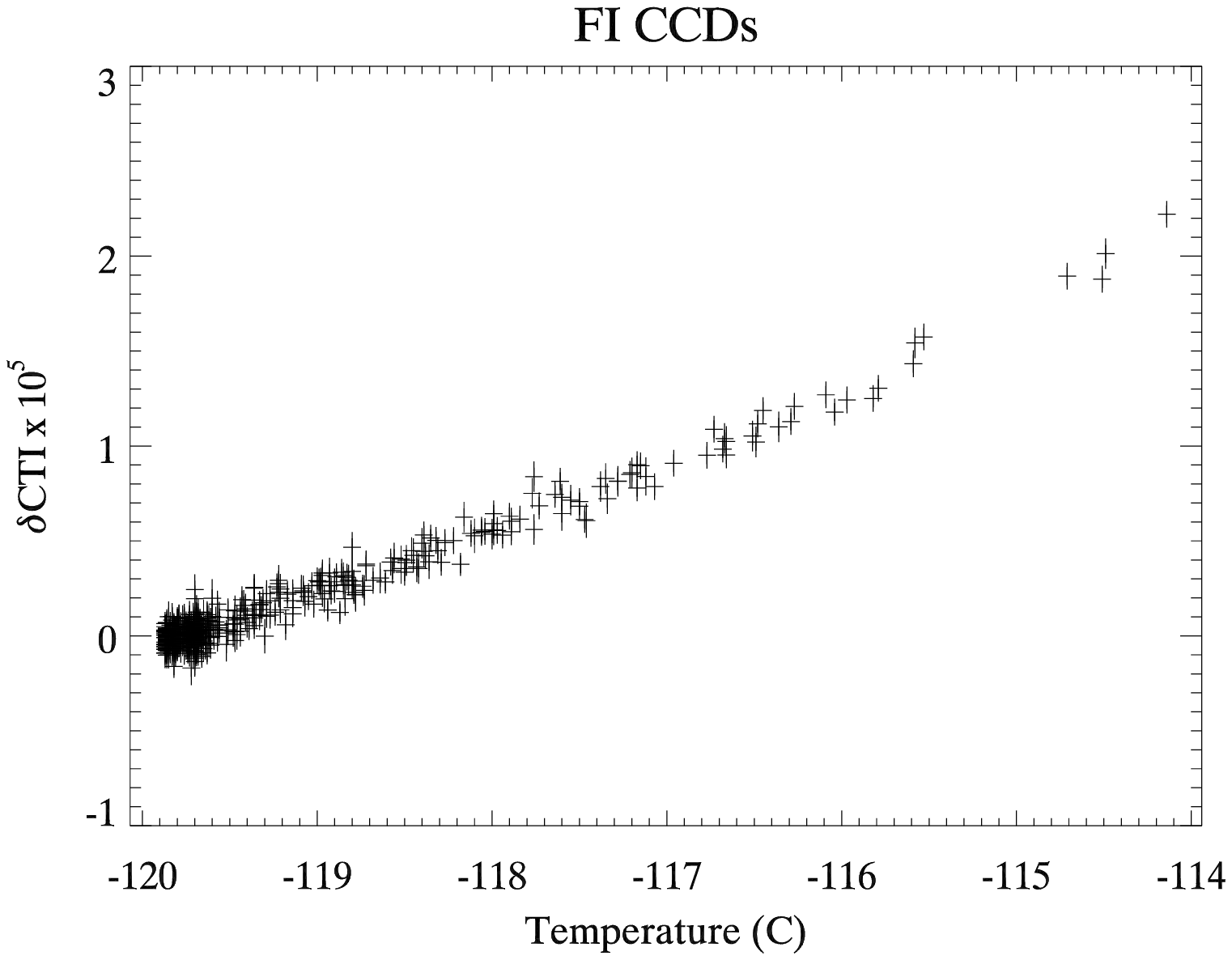}
\includegraphics{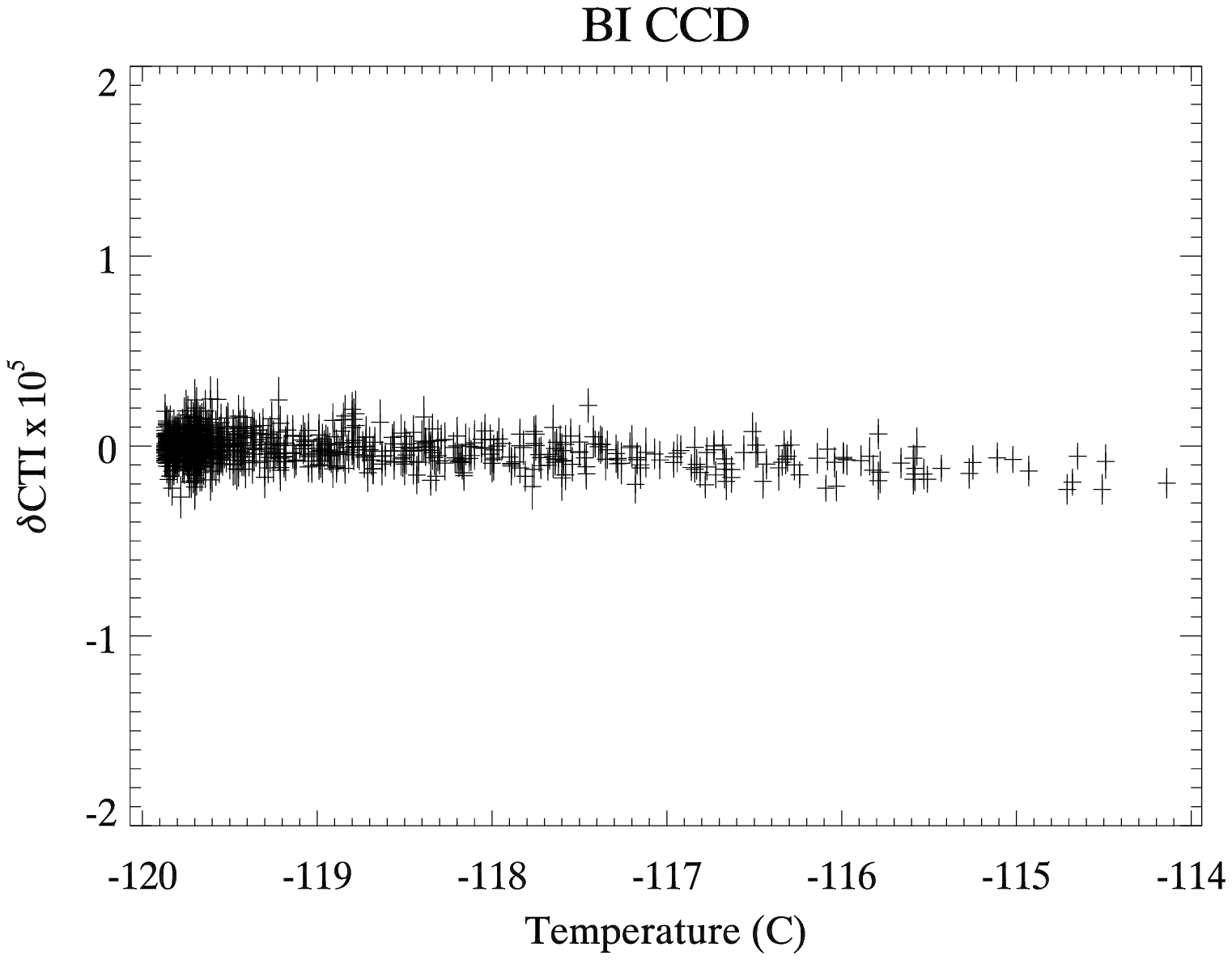}
\caption{The temperature dependence of parallel CTI for the I-array FI CCDs (left) and the ACIS-S3 BI CCD (right).  The particle background variations and the secular CTI increase have been removed.}
\label{fig:ctitemp}
\end{figure}

Since the time constants of the traps that cause CTI are dependent on temperature, we expect some variation of measured CTI due to focal plane temperature deviations.  To monitor the level of radiation damage, these temperature effects must be removed, either by filtering the data or by correcting it.  Figure~\ref{fig:ctitemp} shows the temperature dependence of CTI for the I-array FI CCDs and the ACIS-S3 BI CCD after removing the particle background variations and the secular CTI increase.  The FI CCD CTI is strongly correlated with temperature while the BI CTI has a weak anti-correlation.  The serial CTI on the BI CCDs is weakly correlated with temperature.  An increase in temperature of one degree produces an increase in FI CCD CTI of $\sim 3.4 \times 10^{-6}$ (2.2\%), a decrease of $\sim 2.4 \times 10^{-7}$ (1.3\%) in BI parallel CTI and an increase of $\sim 4.3 \times 10^{-7}$ (0.5\%) in BI serial CTI.  These differences are due to the different nature of the two types of damage.  The FI CTI-temperature relation is best characterized by a second order polynomial, while a linear fit is sufficient for the BI CCD.  We performed the same fit to the CTI-temperature relation for each six month period and found no evidence for time-dependence in this relation.

\subsection{CTI Increase}

\begin{figure}
\vspace{2.6in}
\includegraphics{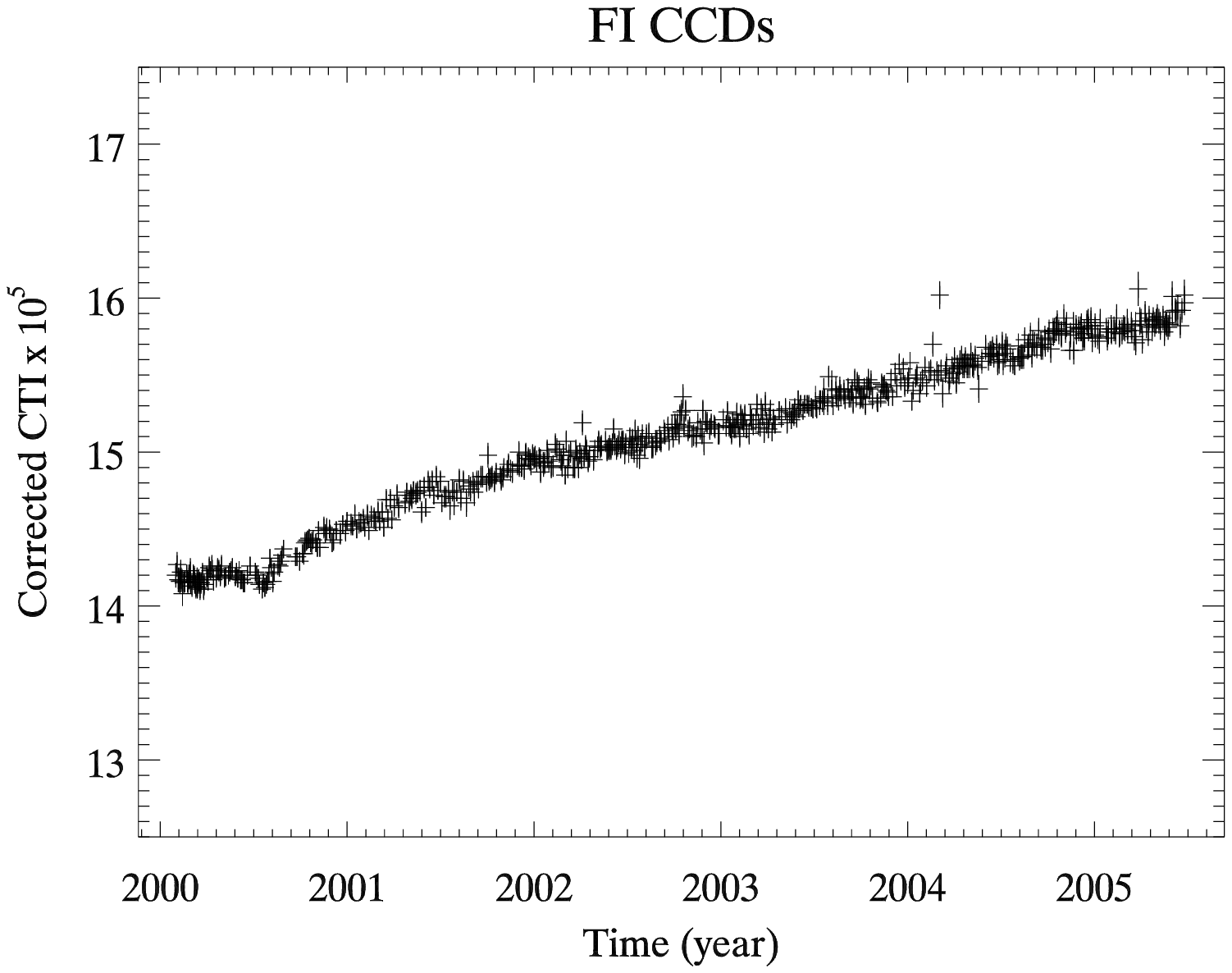}
\includegraphics{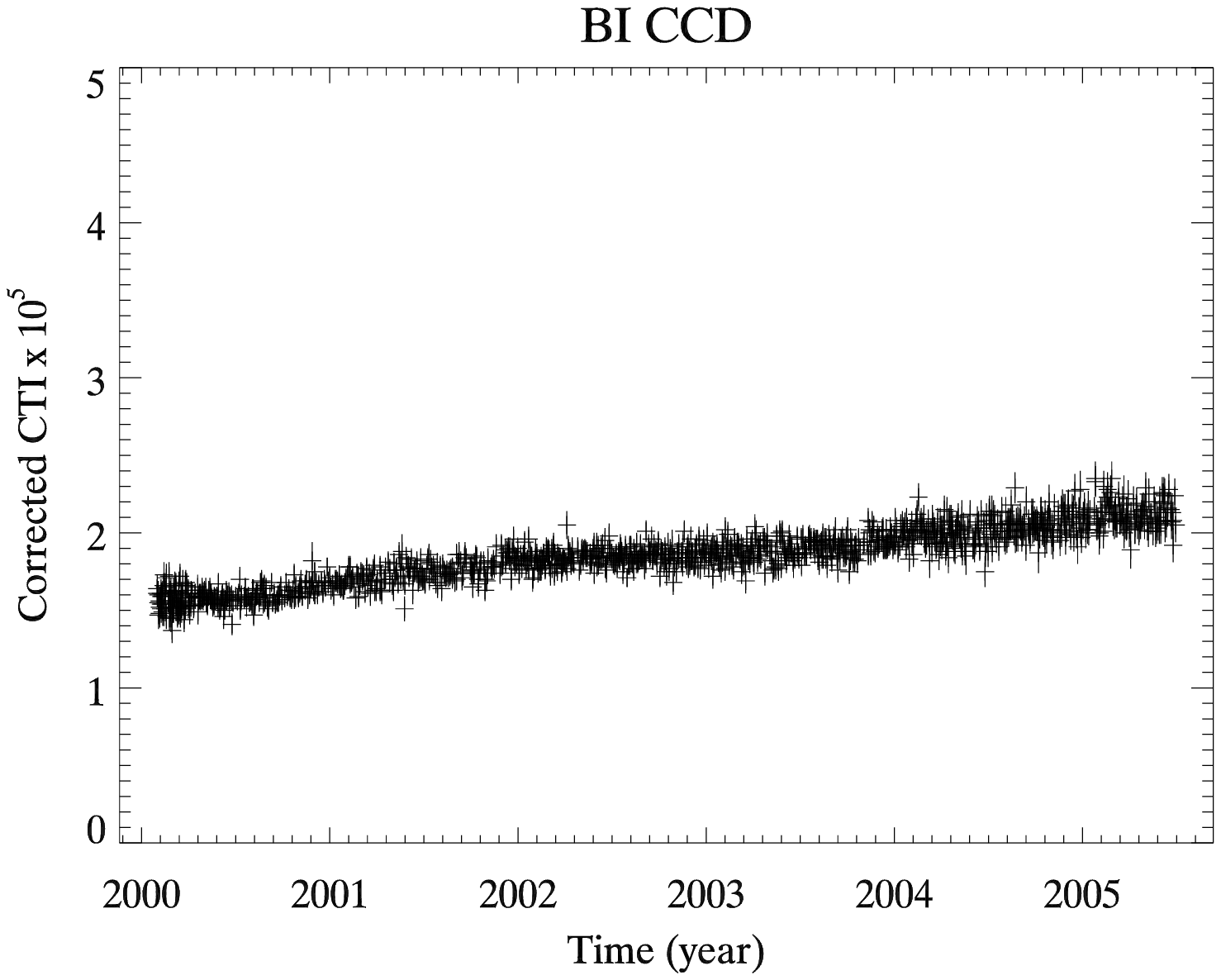}
\caption{Parallel CTI as a function of time for the I-array FI CCDs and the ACIS-S3 BI CCD after correcting for sacrificial charge from the particle background and for temperature variations.  This figure can be compared to the measured uncorrected CTI in Figure~\ref{fig:rawcti}.}
\label{fig:ctifinal}
\end{figure}

\begin{figure}
\vspace{2.7in}
\includegraphics{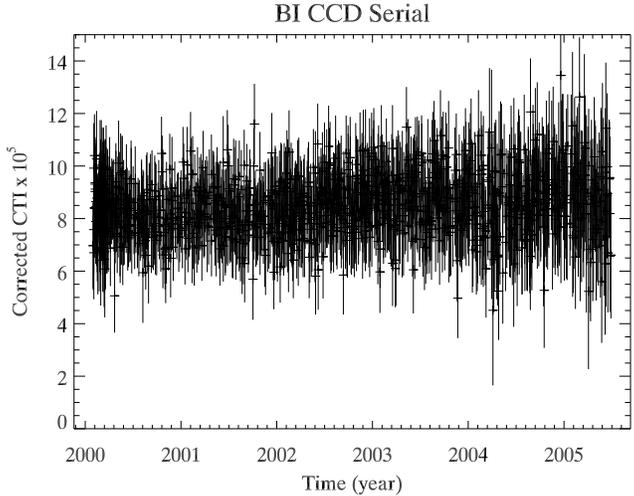}
\caption{Serial CTI as a function of time for the S3 BI CCD after correcting for sacrificial charge and temperature.}
\label{fig:ctiserial}
\end{figure}

After removing the effects due to variations in focal plane temperature and sacrificial charge from the particle background, we can now study the time evolution of radiation damage with the corrected CTI.  Figure~\ref{fig:ctifinal} shows the corrected parallel CTI for the I-array FI CCDs and the ACIS-S3 BI CCD as a function of time.  This can be compared to Figure~\ref{fig:rawcti} which shows the same data without any corrections.  The FI and BI CTI increase now have similar approximately linear shapes.  There is no indication of step-function increases due to specific solar events.  As an example, the ``Halloween'' 2003 solar storm, considered one of the most powerful in this solar cycle, produced no corresponding change in CTI to $< 6 \times 10^{-7}$.  Measuring serial CTI on the FI CCDs is difficult because of the strong parallel CTI component, but the FI CCDs do not yet have any measurable serial CTI.  Figure~\ref{fig:ctiserial} shows the corrected serial CTI for the S3 BI CCD as a function of time.  Neither type of CCD shows evidence for increasing serial CTI.  The change in framestore CTI is measured indirectly and is therefore more uncertain, but a small change in framestore CTI is possibly seen for both types of CCD.  Table~\ref{tab:ctiincrease} lists the initial CTI and the rate of increase after the corrections for temperature and particle background

\begin{table}
\caption{CTI Increase Summary}
\label{tab:ctiincrease}
\begin{center}
\begin{tabular}{lccc}
&Initial CTI &Yearly Increase  &Yearly Increase\\
&($10^{-5}$) &($10^{-6}$ / yr) &( \% / yr) \\
\hline \\
FI Parallel CTI    &$14.17 \pm 0.04$ &$3.29 \pm 0.02$ &$2.18 \pm 0.01$ \\
FI Framestore CTI$^*$ &$\cdots$         &$0.7 \pm 0.1$ &$\cdots$ \\
FI Serial CTI     &$< 2$            &$< 3.6$ &$\cdots$ \\
\\
BI Parallel CTI   &$1.57 \pm 0.07$  &$1.04 \pm 0.02$ &$5.61 \pm 0.08$ \\
BI Framestore CTI$^*$ &$\cdots$         &$1.1 \pm 0.1$   &$\cdots$ \\
BI Serial CTI     &$8.40 \pm 0.98$  &$< 2.3$ &$< 2.7$ \\

\end{tabular}

\smallskip

\noindent
\small
$^*$ Assuming no change in electronic gain. \\
\normalsize
\end{center}
\end{table}

\subsection{Trailing Charge Evolution}

\begin{figure}
\vspace{2.7in}
\includegraphics{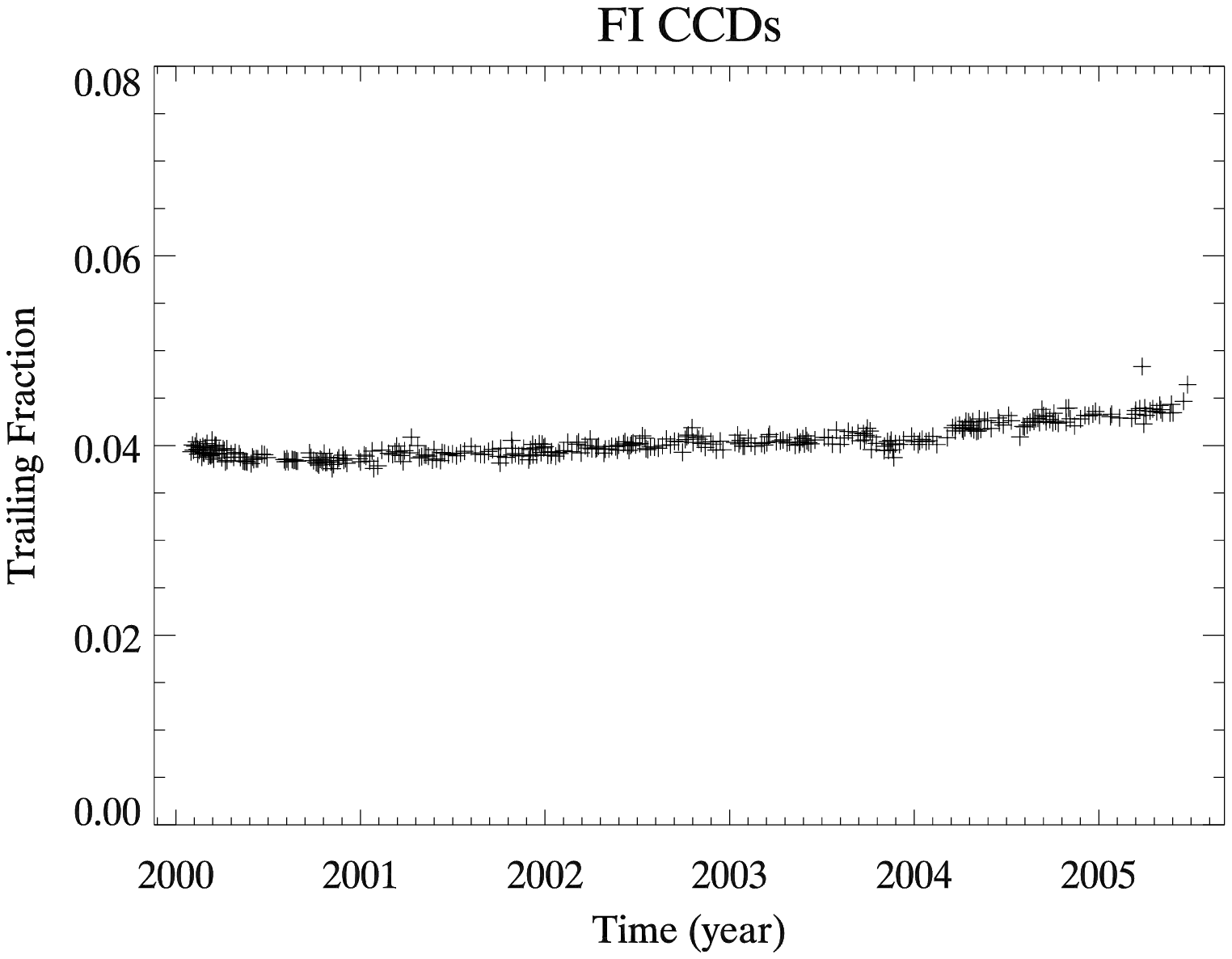}
\includegraphics{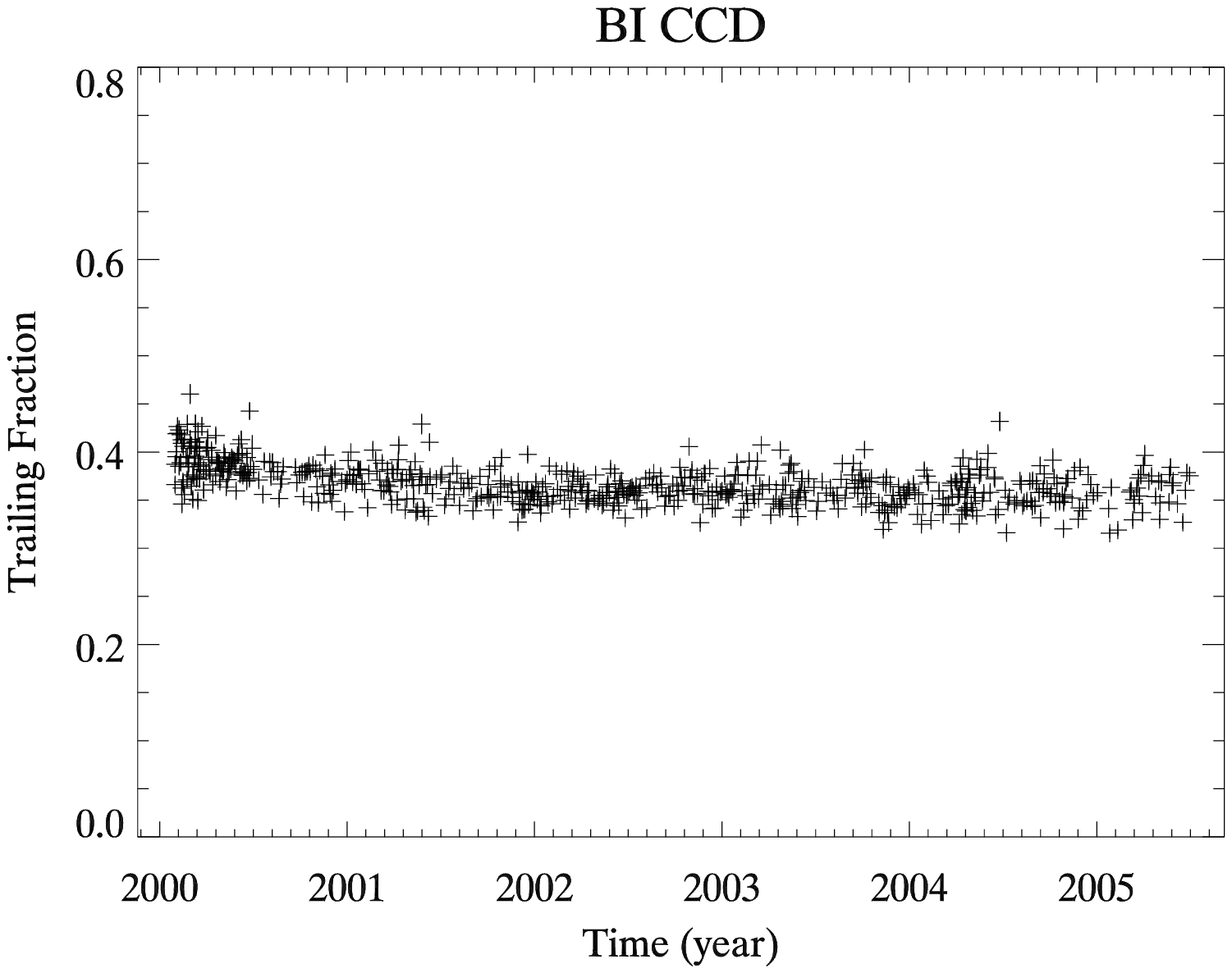}
\caption{The fraction of charge loss emitting into the trailing pixel for the parallel transfer as a function of time.  The I-array FI CCDs is on the left and the S3 BI CCD is on the right.}
\label{fig:trail}
\end{figure}

As described in section~\ref{sec:trail}, the fraction of the charge lost from a pixel that is re-emitted into the next pixel can be used to roughly characterize the time constants of the traps.  Figure~\ref{fig:trail} shows the trailing fraction as a function of time for the I-array FI CCDs and the ACIS-S3 BI CCD.  Since we expect the time constants to be temperature sensitive, we have filtered the data by temperature.  Initially the BI CCDs had a much larger trailing fraction than the FI CCDs, presumably because the BI CCDs have more traps with shorter time constants.  The trailing fraction for each type of CCD has evolved slightly over the past five years.  The FI CCD trailing fraction has increased by $\sim$ 2.1\%, while the BI CCD has decreased by 7.0\%.  Neither the increase nor the decrease appear strictly linear.

\section{Conclusions}
\label{sect:conc}

The data on the increasing radiation damage can be used to speculate as to the nature of the particles causing the damage.  The initial damage was limited to the imaging array of the FI CCDs which implied that the $\sim$~2.5~mm of aluminum in the framestore cover was sufficient to block the damaging particles.  Furthermore, the 40~$\mu$m substrate of the BI CCDs was also sufficient to prevent the damaging particles from reaching the transfer channel.  These characteristics, plus modelling of the environment and mirrors, led to the conclusion that low energy protons (few hundred keV) scattered off the Chandra mirrors were the likely culprit.  

The subsequent damage does not appear to be identical to the initial damage.  Unlike the initial damage, the parallel CTI of the BI CCDs has increased and there is some evidence for an increase in CTI in the framestore array.  This implies that some of the damage is from higher energy particles ($>$ 1 MeV).  Low energy protons must still be present however, since the increase in parallel CTI of the FI CCDs is more than three times that of the BI CCDs.  The upper limits on changing serial CTI do not put any strong restrictions on the nature of the damaging particles.

In addition to the CTI evidence, the evolution of the trailing fraction implies that the long-term radiation damage is not produced by the same spectrum of particles that produced the initial damage.  The FI CCDs show a small increase in the trailing fraction which implies that there are relatively more short time constant traps than initially, while the BI CCDs which started with a much higher proportion of short time constant traps than the FI CCDs, are trending towards longer traps.

Since low energy protons cannot reach the ACIS CCDs except when the detector is in the telescope focal plane, and ACIS not in the focal plane during transits through Earth's radiation belts, the portion of the damage from low energy protons must occur during the remaining part of the orbit when Chandra is observing astrophysical sources.  In contrast, the high energy particles, depending on the energy, may be penetrating the spacecraft and so can cause damage during any part of the orbit.

ACIS radiation management strategy stows ACIS during radiation belt passages and during times of high solar or geomagnetic activity\cite{odell}.  These safety measures can reduce spacecraft observing efficiency, however the slow rate of radiation damage increase can be taken as validation that these measures have been effective.  In particular, there is no indication of step-function increases due to particular solar storms even though our measurements include times of extremely high solar activity such as the ``Halloween'' 2003 event.  While our apparent success at preventing such damage is gratifying, we plan to continue careful monitoring of ACIS performance for the lifetime of the instrument - hopefully a very long time.

\acknowledgments

We would like to thank Paul Plucinsky and the Chandra Science Operations team and Steve O'Dell and the Project Science team for many years of fruitful collaboration in understanding and managing ACIS radiation damage.  This work was supported by NASA contracts NAS 8-37716 and NAS 8-38252.


\bibliography{paper}
\bibliographystyle{spiebib}

\end{document}